\documentclass[aps,prl,twocolumn,floats,showpacs]{revtex4}
\usepackage{graphicx}
\usepackage{amsmath}
\usepackage{amssymb}
\usepackage{units}
\usepackage[usenames,dvipsnames]{color}

\renewcommand{\a}{{\alpha}}

\newcommand{\w}{{\omega}}

\newcommand{\E}{{\cal E}}
\newcommand{\W}{{\Omega}}
\newcommand{\D}{{\Delta}}

\newcommand{\F}{{\cal F}}

\newcommand{\beq}{\begin{equation}}
\newcommand{\eeq}{\end{equation}}
\newcommand{\bea}{\begin{eqnarray}}
\newcommand{\eea}{\end{eqnarray}}

\begin{document}
\title{Coherent Ro-vibrational Revivals in a Thermal Molecular Ensemble}

\date{\today}
\author{M.~Bitter$^{1}$, E.~A.~Shapiro$^{2}$  and V.~ Milner$^{1}$}
\affiliation{Department of  Physics \& Astronomy and The
Laboratory for Advanced Spectroscopy and Imaging Research
(LASIR)$^{1}$, and Department of Chemistry$^{2}$, The University
of British Columbia, Vancouver, Canada \\}

%------------------------Abstract---------------------------------------------
\begin{abstract}{
We report an experimental and theoretical study of the evolution of vibrational coherence in a thermal ensemble of nitrogen molecules. Rotational dephasing and rephasing of the vibrational coherence is detected by coherent anti-Stokes Raman scattering. The existence of ro-vibrational coupling and the discrete energy spectrum of the rotational bath lead to a whole new class of full and fractional ro-vibrational revivals. Following the rich ro-vibrational dynamics on a nanosecond time scale with sub-picosecond time resolution enables us to determine the second-order ro-vibrational constant $\gamma_e$ and assess new possibilities of controlling decoherence.}
\end{abstract}


\maketitle

\section{Introduction}
%--------- Introduction 1: Coupling to environment, dephasing as decoherence, rephasing as revivals-------
Coupling of a quantum system to an external environment leads to a loss of coherence between the eigenstates of the system, and therefore to a loss of information \cite{Bartram2010,Brif2001, Shapiro2003,Joos-book,BP-book}. For all processes that rely on coherent dynamics, it is of general interest to understand the evolution of an open system and to minimize the decoherence effects (the general topic of decoherence is covered in a number of recent reviews \cite{Zurek2002, Shapiro2003}). If the bath associated with the environment is large and its spectrum is dense, the information about the system of interest can no longer be recovered, and only in some cases the decoherence can be slowed down \cite{Katz2007,Shapiro2007,Brif2001,Branderhorst2008,NC-book}. Coupling to a small bath with a spectrum consisting of only a few lines can lead to recurrences in the information flow. The discrete spectrum leads to a periodic dephasing and rephasing of the constituent eigenstates.

In general, \textit{dephasing} describes the loss of coherence, whereas \textit{rephasing} results in quantum revivals when the system approaches its original state. A partial recovery of the initial state, in which some of the excited eigenstates undergo a complete rephasing while others do not, is known as a fractional revival \cite{Averbukh1989}.

%------Introduction 2: Ro-vibrational coupling as a bath - Ro-vibrational decoherence and revivals
Diatomic molecules represent one of the most studied quantum systems exhibiting revivals and decoherence due to the molecular rotation. Extensive studies have been carried out on the rotational revivals \cite{Seideman1999,Rosca-Pruna2002,Comstock2003}, as well as the dephasing of the molecular vibration due to the ro-vibrational coupling to the bath of many thermally populated rotational states \cite{Bartram2010, Wallentowitz2002, Branderhorst2008}. The mathematical description of the ro-vibrational dephasing on a short time scale is that of decoherence \cite{Wallentowitz2002, Brif2001, Adelswaerd2004}. Partial rephasing of the ro-vibrational dynamics has been observed experimentally in molecular iodine \cite{Dantus1990}. On a longer time scale, full rephasings of the excited ro-vibrational eigenstates (hereafter referred to as ``ro-vibrational revivals'') have been predicted \cite{Wallentowitz2002, Hansson2000}, yet their observation has been considered unlikely due to the difference between the time scales of the vibrational, rotational, and ro-vibrational dynamics \cite{Wallentowitz2002}.

The meaning of a ro-vibrational revival is quite different from its rotational analogue. The latter leads to the periodic reappearance of molecular alignment when the internuclear axes of the molecules rotating with different angular velocities coincide in space \cite{Seideman1999}. In contrast, the ro-vibrational revival happens when all rotating molecules vibrate in phase with one another regardless of their spatial orientation.

%------------------Introduction 3: Goal of this work ------------------
In this paper, we study the evolution of a vibrational coherence in an ensemble of nitrogen molecules. The initial coherence between $v=0$ and $v=1$ vibrational levels is created with a two-photon Raman excitation. Due to the presence of the thermal bath of rotational levels and ro-vibrational coupling, the vibrational coherence is lost on a short (few picosecond) time scale. This process has been observed and described theoretically  in a number of papers, and served as a prototype mechanism for studying decoherence \cite{Bartram2010, Branderhorst2008, Shapiro2007}. We follow the evolution of the vibrational coherence on a longer (nanosecond) time scale with femtosecond resolution, and measure it directly by means of coherent anti-Stokes Raman scattering (CARS). As a consequence of the discreteness and the relatively low density of rotational states of nitrogen, we are able to observe full and fractional revivals of the vibrational coherence. Our theoretical analysis explains the observed dynamics and allows accurate retrieval of the ro-vibrational constants of N$_2$ from the experimental data.

%-----------Introduction 4: Motivation: Coherent control of decoherence---------------------
Controllability of molecular dynamics in the presence of coupling to the environment has been a subject of thorough study and debate \cite{Bartram2010,Branderhorst2008,Shapiro2007}. The ability to control decoherence is important in the field of quantum information and quantum computation \cite{Lidar1998,Palma1996,Shapiro2003}. Different approaches to controlling decoherence have been suggested, including phase-only control \cite{Bartram2010}, feedback control \cite{Katz2007}, generation of decoherence free subspaces \cite{Lidar1998} and suppression of decoherence by means of nonlinear resonances \cite{Shapiro2007}. Methods of adaptive coherent control \cite{Judson1992} have been implemented to inhibit dephasing in small molecules \cite{Branderhorst2008} and molecules as complex as proteins \cite{Prokhorenko2006}. Our experimental and theoretical study of the molecular ro-vibrational dynamics, presented here, aims at addressing the fundamental problem of decoherence control in systems coupled to the environment.

\section{Theory}
%-------------- Theory 1: Ro-vibrational spectrum  --------------------------------------------
We start by considering the coherence between two vibrational states $v_0$ and $v_1$ of a diatomic molecule. Using the Dunham expansion, the energy of an eigenstate with vibrational number $v$ and rotational number $J$ can be written as:
\begin{equation}
\E(v,J) = hc\left[G(v)+F(v,J)\right], \label{eq-EnergyGeneral}
\end{equation}
where $h$ is the Planck's constant, $c$ is the speed of light in vacuum, and 
\begin{align}\label{eq-VibEnergies}
    &G(v) = \omega_e(v+1/2) - \omega_e x_e(v+1/2)^2 + \omega_e y_e(v+1/2)^3, ~~~~\\
    &F(v,J) = B_v J(J+1) - D_v J^2 (J+1)^2,
    \label{eq:Energies}
\end{align}
with $\omega_e$, $\omega_e x_e$, $\omega_e y_e$ being the vibrational constants. The rotational constants $B_v$ and $D_v$ depend on the vibrational quantum number $v$ according to:
\begin{align}
    B_v &= B_e - \alpha_e(v+1/2) + \gamma_e(v+1/2)^2, \\
    D_v &= D_e + \beta_e(v+1/2).
    \label{eq:B}
\end{align}
The term $\beta_e$ can generally be neglected on the relevant timescale of 1 ns. The spectroscopic constants for nitrogen, studied in this work, are given in Table \ref{tab:ConstantsN2}.

\begin{table*}
\begin{center}
    \begin{minipage}{1\linewidth}
            \begin{tabular}{ p{1cm} | p{1.4cm} p{1.4cm} p{1.4cm} p{2.2cm} p{1.4cm} p{2.2cm} p{2cm} p{2cm} p{1.0cm}}
        & $T_e$ &    $\omega_e$ & $\omega_e x_e$ & $\omega_e y_e$   &   $B_e$       & $\alpha_e$ &              $\gamma_e$ & $D_e$ & $\beta_e$ \\
        \hline\hline
X$^1\Sigma_g^+$     & 0     &           2358.57 &       14.324  &           $-2.26\times 10^{-3}$                   & 1.99824   & $1.7318\times 10^{-2}$    & \textit{NA}  & $5.76\times 10^{-6}$ & \textit{NA}\\
A$^3\Sigma_u^+$     & 50203.6 & 1460.64  &  13.87  &   0.0103   &  1.4546  &  0.018 & $-8.8\times 10^{-5}$ & $6.15\times 10^{-6}$ & \textit{NA} \\
            \end{tabular}
            \caption{Vibrational and rotational constants for the electronic ground state and the first electronic excited state of $^{14}$N$_2$. \textit{NA} indicates that this value is not available on the NIST database \cite{NIST}. All values are in cm$^{-1}$. }
            \label{tab:ConstantsN2}
    \end{minipage}\hfill
\end{center}
\end{table*}

%-------------- Theory 2: Time evolution --------------------------------------------
The evolution of the molecular ensemble in our experiment can be divided into three stages: excitation, field-free dynamics, and probing. We begin with an ensemble of molecules in the ground electronic and vibrational state. The initial rotational distribution is thermal, and is therefore given by the Boltzmann weights
\begin{equation}
    P_J = \frac{hcB_0}{kT}(2J+1) \exp\left(- \frac{hcB_0}{kT}J(J+1)   \right)
\label{eq:Boltzman}
\end{equation}
with the Boltzmann constant $k$ and temperature $T$.

At the first stage of the evolution, pump and Stokes laser pulses excite coherence between $v=0$ and $v=v_1$ vibrational levels (Fig.\ref{Fig:CARSscheme}(a)). We assume that both fields are linearly polarized and of moderate strength, and are therefore driving only transitions $(v=0,J,M_J) \rightarrow (v=v_1, J'=\{J,J\pm2\}, M'_{J})$. The process of excitation is considered in details in the Appendix. Here, we assume that the excitation amplitude for all transitions with $J'=J$ ($Q$ branch) is equal to $C_Q$, and for all $J'=J\pm2$ transitions ($O$ and $S$ branches) to $C_{OS}$.

At the second stage, the wave function of each $v,J,M_J$ state acquires the phase
\begin{equation}
\phi(v,J,M_J) = -i\,\E(v,J)\,t\,/\hbar
 \label{eq-phase}
\end{equation}
with energy $\E(v,J)$ given in Eqs.(\ref{eq-EnergyGeneral}-\ref{eq:B}). The coherence between $v=0$ and $v=v_1$ levels in the vibrational density matrix evolves as
\begin{equation}
\rho_{01}(t) =  \F_{Q}(t) + \F_{OS}(t)%
\label{eq-coherence-main}
\end{equation}
where, assuming that $|C_Q|$ and $|C_{OS}|$ are small compared to
1 (perturbative regime),
\begin{eqnarray}
\F_Q(t) &=& e^{-i \w_{01}t}\,C_Q  \sum_J P_J e^{- i 2\pi c
(B_{v1} -
B_{0})J(J+1) t},  \label{eq-coherence-time-evolution-Q} \\
\F_{OS}(t) &=& e^{-i \w_{01}t}\, C_{OS}  \label{eq-coherence-time-evolution-PS} \\%
 && \times\,\sum_{J,\pm} P_J e^{- i 2\pi c
[B_{v1} (J\pm2)(J+1\pm2) - B_{0} J(J+1) ]t}, \nonumber
\end{eqnarray}
with $\w_{01} = {2\pi c} [G(v_1)-G(0)]$.

At the third stage, the probe pulse interacts with the molecules, generating anti-Stokes radiation. The latter arises due to the
Raman coherences between the original $v=0,J,M_J$ eigenstates and the excited $v=v_1,J',M'_{J}$ states populated by the two-photon pump-Stokes field. Coherences $\F_Q$, responsible for the $Q$-branch scattering, and $\F_{OS}$ ($O$ and $S$ branches) exhibit different temporal dynamics.

%-------------------Theory 3: Ro-vibrational revivals -----------------------------------
The evolution of  $\F_Q(t)$ has been studied theoretically in detail in Refs.\cite{Wallentowitz2002,Bartram2010}. In many aspects it is similar to that of a conventional quantum wave packet \cite{Averbukh1989,Schleich-book} with state amplitudes given by the coefficients $P_J$, and the (effective) energies
\begin{equation}
\E_\text{eff}=hc(B_{v_1}-B_{0})J(J+1)  ~. \label{eq-RovibSpectrum}
\end{equation}
Similarly to the evolution of a pure rotational coherence, the ro-vibrational evolution is governed by the interference between the molecules occupying different angular momenta states, whose relative phase evolves at frequencies proportional to $J(J+1)$. However, in contrast to the relative rotational phase between the constituents of a purely rotational wave packet, it is the relative phase between the vibrational coherence $\rho _{v,v_1}$ in molecules occupying different $J$ states that dictates the ro-vibrational dynamics considered in this work (Fig.\ref{Fig:CARSscheme}(b)). Coherent ro-vibrational evolution is observed in a thermal mixture of rotational states rather than in a coherent superposition.

As follows from Eqs.(\ref{eq-coherence-time-evolution-Q},\ref{eq-RovibSpectrum}) and by direct analogy with rotational dynamics, the evolution of $\F_Q$ is defined by the ``ro-vibrational revival" time
\begin{equation}
    T_\text{RoVib} =  \frac{1}{2c |B_{v1}-B_{0}|}~.
\end{equation}
This value can be varied by choosing an appropriate excited vibrational state $v_1$. In our experiment, $v_1=1$, and
\begin{equation}
    T_\text{RoVib} =  \frac{1}{2c(\alpha_e -2\gamma_e)}~.
\label{eq-Trovib}
\end{equation}
After pump and Stokes pulses create the rotational coherence, different rotational states in the sum of Eq.(\ref{eq-coherence-time-evolution-Q}) begin to dephase, and the amplitude of $\F_Q$ decreases. The duration of the initial dephasing period is defined by the difference in the effective energy between the states with low and high values of $J$:
\begin{equation}
T_\text{dephasing} \sim \frac{T_\text{RoVib} }{\D J^2},
\label{eq-Tdephasing}
\end{equation}
where $\D J$ is set by the thermal distribution (\ref{eq:Boltzman}). During this period of initial dephasing, the discrete nature of the rotational bath is not resolved, and the process is equivalent to the conventional decoherence \cite{Bartram2010,Wallentowitz2002}. Later, at $t=T_\text{RoVib}$, the coherence undergoes a full revival: each term in the sum of Eq.(\ref{eq-coherence-time-evolution-Q}) acquires a phase equal to integer multiple of $2\pi$. At intermediate times, $\F_Q$ undergoes a series of fractional revivals whose structure  has been theoretically described in Ref.\cite{Wallentowitz2002}.

The second part of the ro-vibrational coherence, $\F_{OS}(t)$, is due to the  Raman excitations with $\Delta J=\pm2$. Its dynamics are revealed on a much faster rotational time scale. Indeed, at short times the difference between $B_{v1}$ and $B_0$ can be neglected, and expression (\ref{eq-coherence-time-evolution-PS}) is similar to that describing the conventional non-adiabatic alignment of molecules \cite{Stapelfeldt2003,Heritage1975}. The wave packet (\ref{eq-coherence-time-evolution-PS}) revives in time at
\begin{equation}
    T_{\text{rot}}= \frac{1}{2cB_0}~.
\end{equation}
At longer times, the shape of these revivals becomes more complex due to the  difference between $B_{v1}$ and $B_0$.

\section{Experiment}
\begin{figure}
\centering
 \includegraphics[width=1.0\columnwidth]{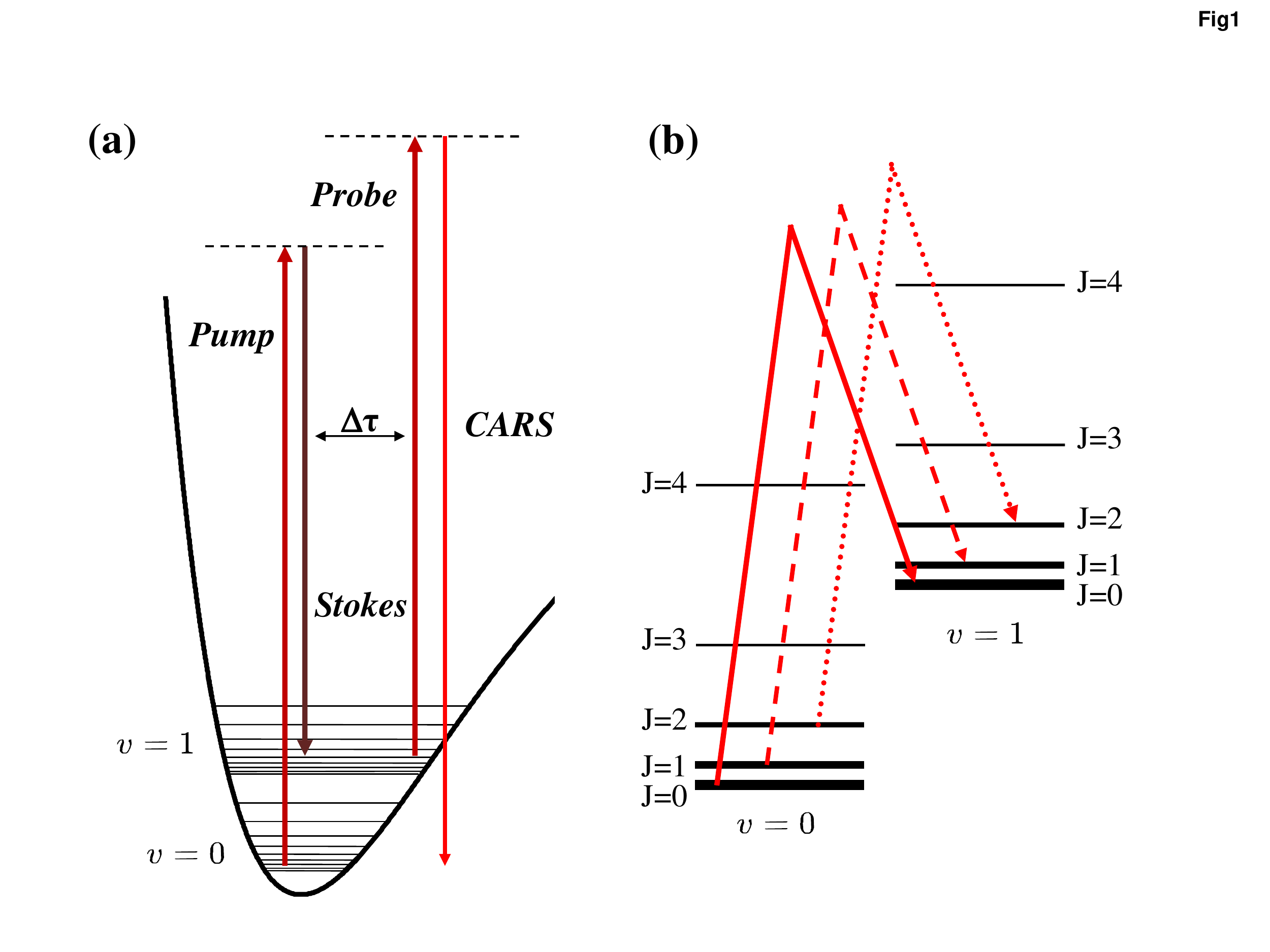}
     \caption{(Color online) (a) Time-resolved CARS on nitrogen molecules: pump and Stokes wavelengths are chosen to match the energy gap between the vibrational states $v=0$ and $v=1$. Probe pulses arrive after a time delay $\Delta \tau$ and generate the CARS signal. (b) Illustration of different $Q$-branch Raman transitions which contribute to the vibrational coherence.}
  \vskip -.1truein
  \label{Fig:CARSscheme}
\end{figure}

%--------------  Experiment 1: CARS -----------------------------------------------
We use coherent anti-Stokes Raman scattering (CARS) to measure vibrational coherence in nitrogen molecules, as shown in Fig.\ref{Fig:CARSscheme}. Two laser fields, pump and Stokes with frequencies $\omega_\text{p}$ and $\omega_\text{S}$, respectively, prepare vibrational coherence. Probe field with frequency $\omega_\text{pr}$ scatters off this coherence, generating anti-Stokes field at $\omega_{aS}=\omega_\text{p}-\omega_\text{S}+\omega_\text{pr}$. The ro-vibrational dynamics is detected by scanning the timing of the femtosecond (fs) probe pulse.

%--------------Experiment 2: Setup-----------------------------------------------
\begin{figure}
\centering
 \includegraphics[width=1\columnwidth]{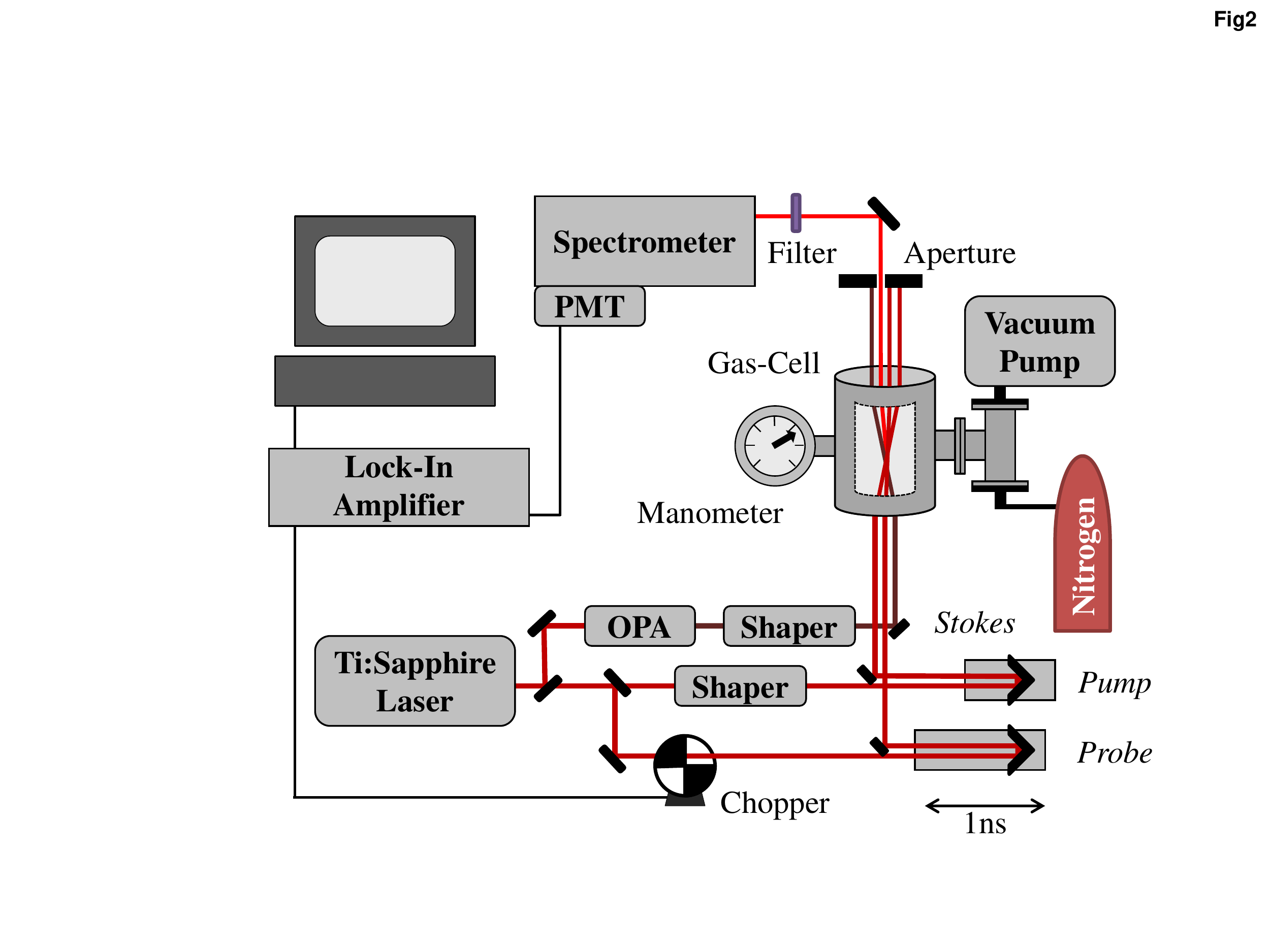}
     \caption{(Color online) Experimental setup. See text for details.}
  \vskip -.1truein
  \label{Fig:Setup}
\end{figure}

Our experimental setup is shown schematically in Fig.\ref{Fig:Setup}. A Ti:Sapphire-based laser system (SpitFire Pro, Spectra-Physics) produces 2 mJ 130 fs pulses at 800 nm and 1 KHz repetition rate. These pulses serve as pump and probe, whereas an optical parametric amplifier (TOPAS, Light Conversion) generates Stokes pulses at a wavelength of 982nm.

All three pulses are spatially overlapped inside a vapor cell filled with nitrogen $^{14}$N$_{2}$ at room temperature under variable pressure. Folded BOXCARS geometry \cite{Shirley1980} provides phase matching between the incident beams, and enables spatial filtering of the output anti-Stokes beam. After passing through spatial and spectral filters, the signal is sent to a spectrometer equipped with a photomultiplier tube detector. To further improve the signal-to-noise ratio, lock-in detection is utilized with the probe beam being optically chopped at a frequency of 360Hz. Pump, Stokes and probe pulses are focused down to a beam diameter of $17\mu$m, $20\mu$m and $12\mu$m, respectively, resulting in peak intensities of $6 \cdot 10^{13}$ W/cm$^2$, $3 \cdot 10^{13}$ W/cm$^2$ and $2 \cdot 10^{12}$ W/cm$^2$, respectively. Two delay lines are used to vary the relative timing of pump, Stokes and probe pulses. Pump and Stokes pulses are shaped by two separate pulse shapers implemented in a standard '$4f$'-geometry \cite{Weiner2000}.

\begin{figure}
\centering
    \includegraphics[width=1.0\columnwidth]{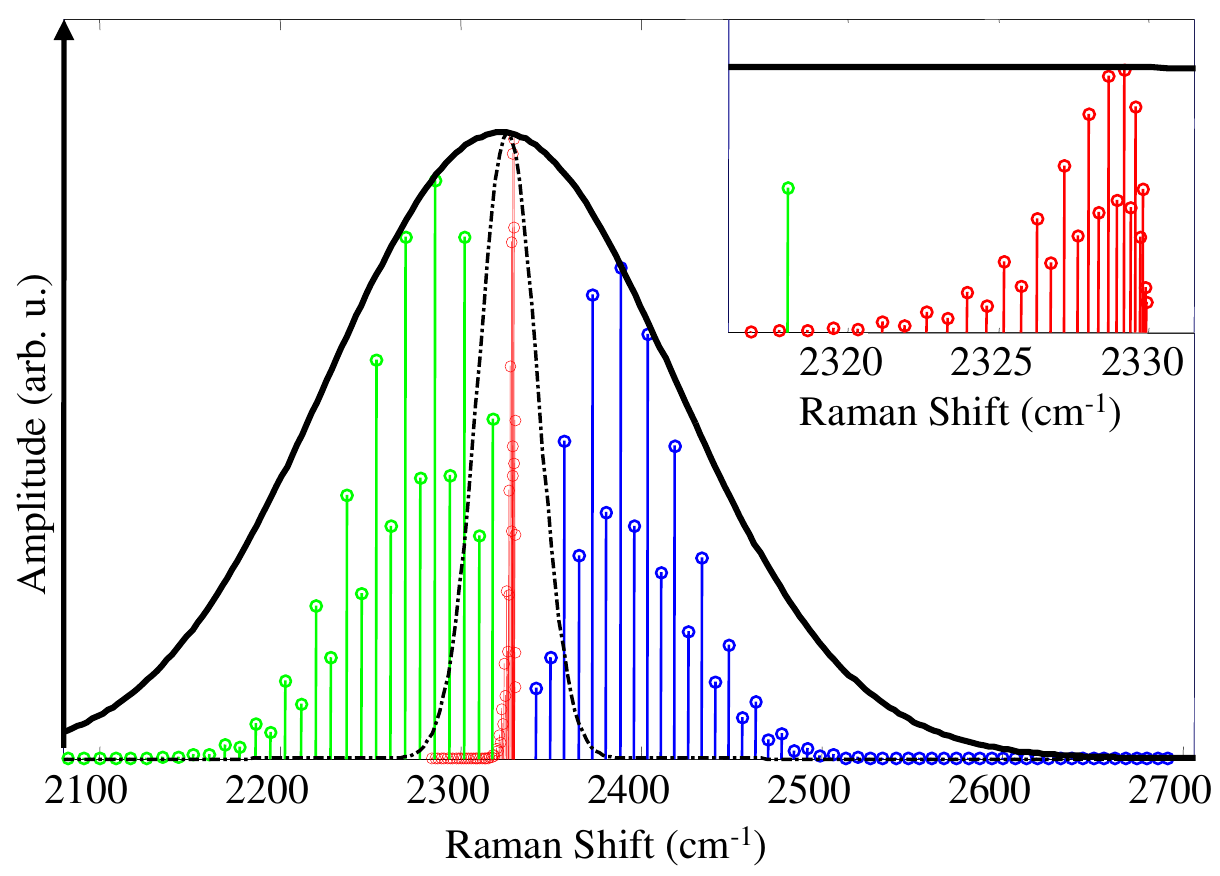}
     \caption{(Color online) Two-Photon spectrum of the pump-Stokes field with transform-limited pulses (solid black line) and after applying a frequency chirp of $\alpha=35,000$fs$^2$ (dashed black line) to both excitation pulses. All possible Raman transitions are indicated: $O$-branch (green), $Q$-branch (red), $S$-branch (blue).  The inset is zoomed in on the $Q$-branch structure. }
  \vskip -.1truein
  \label{Fig:TwoPhotonSpectrum}
\end{figure}

%-------------- TWO PHOTON SPECTRUM --- Choice of Wavelength ---------------------------
The Stokes wavelength of $982$ nm is chosen to match the two-photon spectrum shown in Fig.\ref{Fig:TwoPhotonSpectrum}, corresponding to the vibrational level spacing of $v=0$ and $v=1$ levels in nitrogen. At room temperature, there are $\sim30$ populated rotational states, leading to a multitude of possible Raman transitions. The amplitude of each Raman transition depends not only on the
Boltzmann weight (\ref{eq:Boltzman}) but also on the envelope of the two-photon spectrum and the relative statistical weight of 2:1 between the two spin isomers of nitrogen.

\section{Results}
%------------------ Results: Revival Structure --------------------------------------------
\begin{figure*}
\begin{center}
    \begin{minipage}[t]{1\linewidth}
   \includegraphics[width=1\columnwidth]{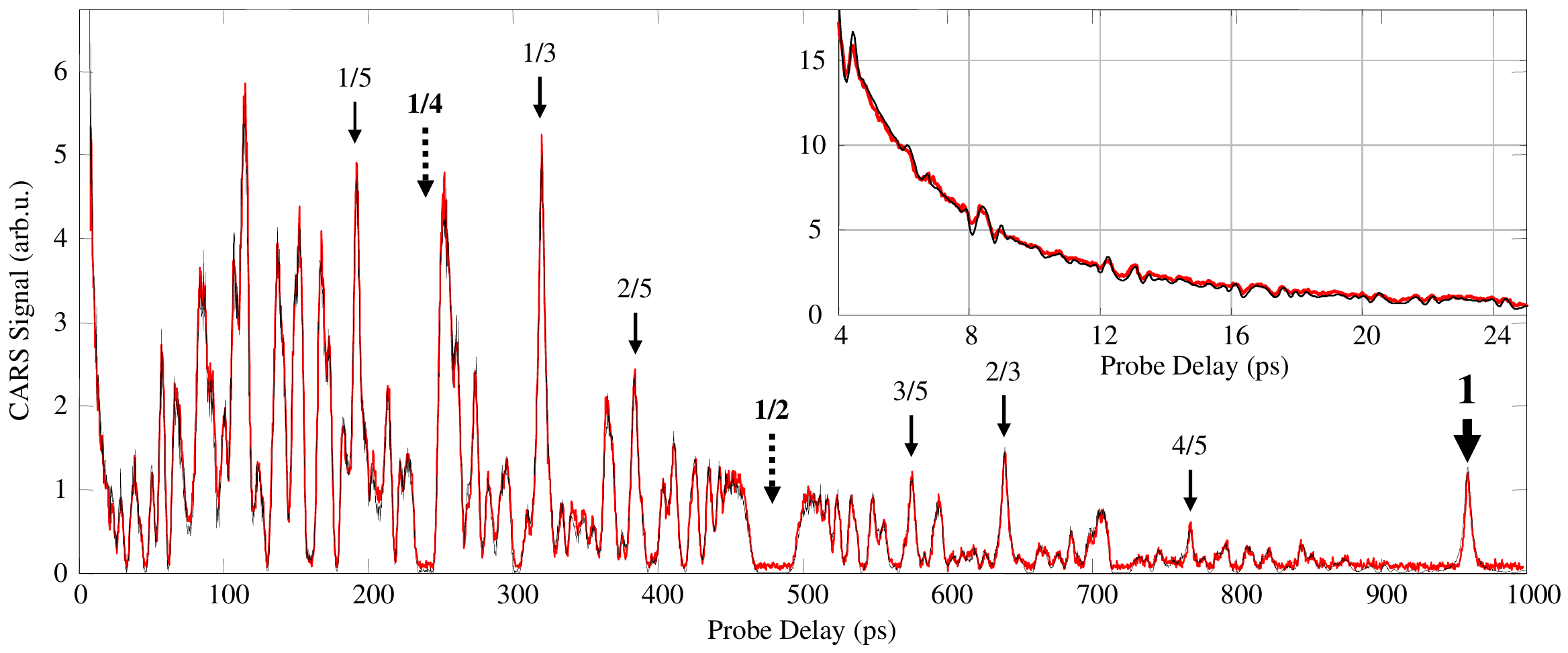}
    \caption{(Color online) Dephasing and rephasing of CARS signal in N$_2$ at room temperature and pressure of $p=158\pm 5$Torr. Experimental data (red) and numerical calculation (black) are almost indistinguishable. The inset shows the dephasing developing on the short time scale of up to 25 ps. }
    \label{Fig:RevivalStructure}
    \end{minipage}\hfill
  \end{center}
\end{figure*}

We investigate the time evolution of the laser-induced ro-vibrational coherence by measuring the magnitude of CARS as a function of the time delay between the excitation and probe pulses. Fig.\ref{Fig:RevivalStructure} shows the experimental results together with the calculated dynamics. For times shorter than 25 ps (inset in Fig.\ref{Fig:RevivalStructure}), the signal is decaying almost monotonically due to the dephasing of the initially prepared wave packet. This time scale is described by Eq.(\ref{eq-Tdephasing}).
Rephasing of the wavepacket at longer times results in a full ro-vibrational revival observed at $T_{\text{RoVib}}=960.2\pm0.3$ ps. On the intermediate time scale, a number of fractional revivals can be clearly seen. They occur at fractions of the full revival time with odd denominators $t=T_{\text{RoVib}} \times \{ \frac{1}{3},\frac{2}{3},\frac{1}{5},\frac{2}{5},\frac{3}{5},\frac{4}{5},... \} $ as predicted by Wallentowitz et.al. \cite{Wallentowitz2002}. The revival peaks manifest the moments in time when the majority of all rotating molecules are vibrating in phase with one another. Similarly, the ``anti-revival'' dips at $\frac{1}{4} T_{\text{RoVib}}$, $\frac{1}{2} T_{\text{RoVib}}$ and $\frac{3}{4} T_{\text{RoVib}}$ correspond to the time instances of out-of-phase molecular vibration.

We simulated the observed signal numerically using the diatomic constants given in Table \ref{tab:ConstantsN2} and the theoretical framework described earlier in the text. Fitting parameters were the collisional decay time and the ro-vibrational constant $\gamma_e$. As can be seen in Fig.\ref{Fig:RevivalStructure}, the calculations are in excellent agreement with the experimental observations.

%------------------  Results: Collisional Decay Time and gamma_e----------------------
At atmospheric pressure the retrieved collisional decay time was $59\pm4$ ps. This decay proved too fast for detecting the full ro-vibrational revival. Lowering the pressure leads to a longer decay time, but also results in a significant drop of the signal strength which scales quadratically with the number of molecules. To observe the revival dynamics, the pressure was set to $158\pm5$ Torr, which resulted in a longer collisional decay time of $256\pm10$ps. The revival dynamics are determined only by those energy terms in Eqs.(\ref{eq-EnergyGeneral}-\ref{eq:Energies}) which depend on both the vibrational quantum number $v$ and rotational quantum number $J$. Using $\gamma_e$ as a second fitting parameter enabled us to determine its value as $\gamma_e=-(2.6\pm0.3)\cdot 10^{-5}$cm$^{-1}$.
We note that the ro-vibrational constant $\beta_e$ has no effect on the fitting accuracy, and can therefore be neglected.

\begin{figure}
\centering
   \includegraphics[width=1.0\columnwidth]{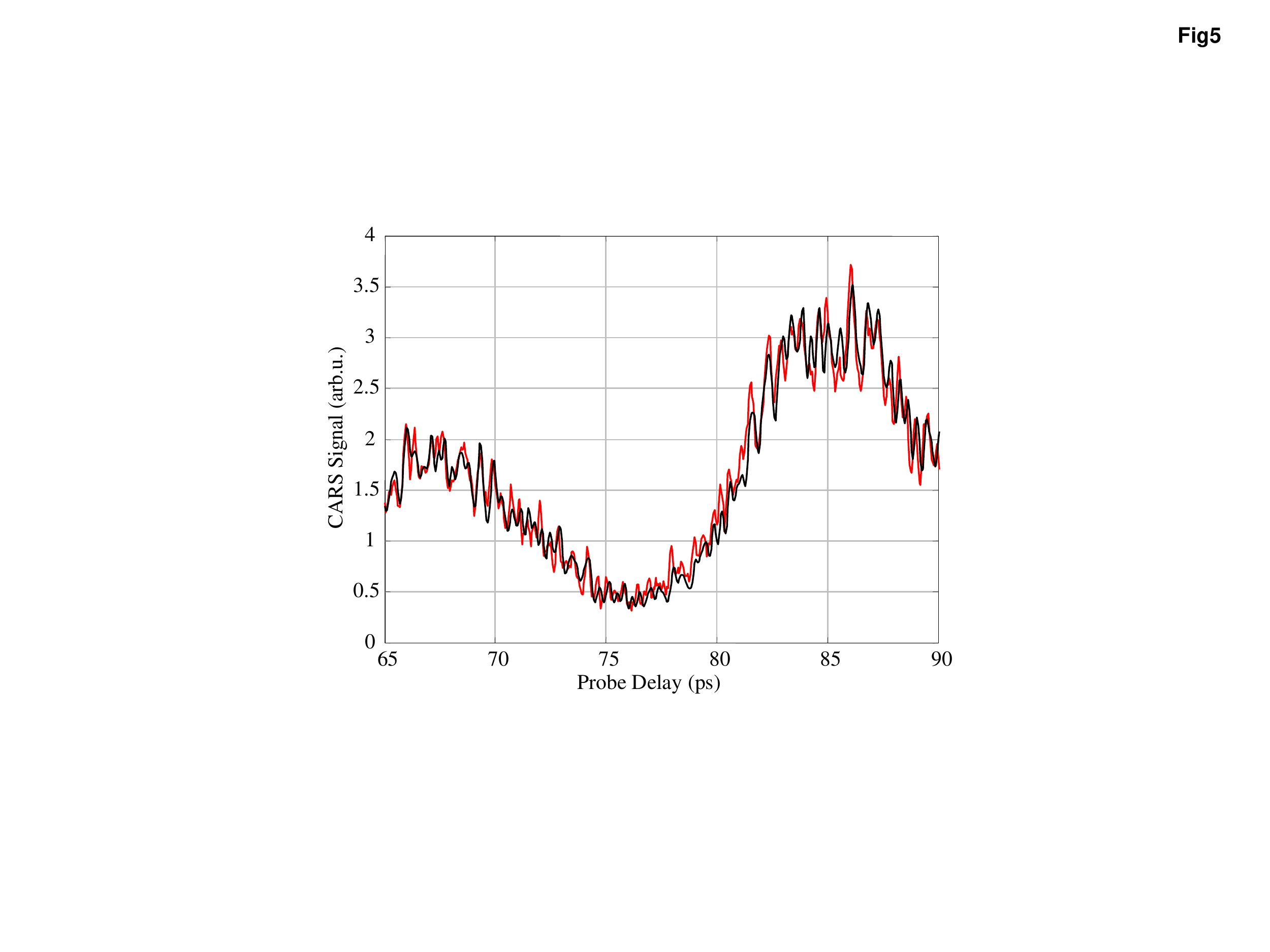}
     \caption{(Color online) Time evolution of CARS signal between $65$ ps and $90$ ps at $P=153\pm3$ Torr:
     experimental data (red), numerical calculation (black).}
  \vskip -.1truein
  \label{Fig:ScanHighResolution}
\end{figure}

%------------------  Results  O & S branch (CHIRPING)---------------------------------
Having a closer look at the rephasing picture reveals a finer structure corresponding to faster ro-vibrational dynamics due to $O$ and $S$ branches of Raman excitation. An example is shown in Fig.\ref{Fig:ScanHighResolution} together with the results of numerical calculations. Note that fast oscillations are of much lower amplitude, in agreement with the ratio of 1:21:1 between the anticipated response for $O$, $Q$ and $S$ branches, respectively (see Appendix for details).

\begin{figure}
\centering
    \includegraphics[width=1.0\columnwidth]{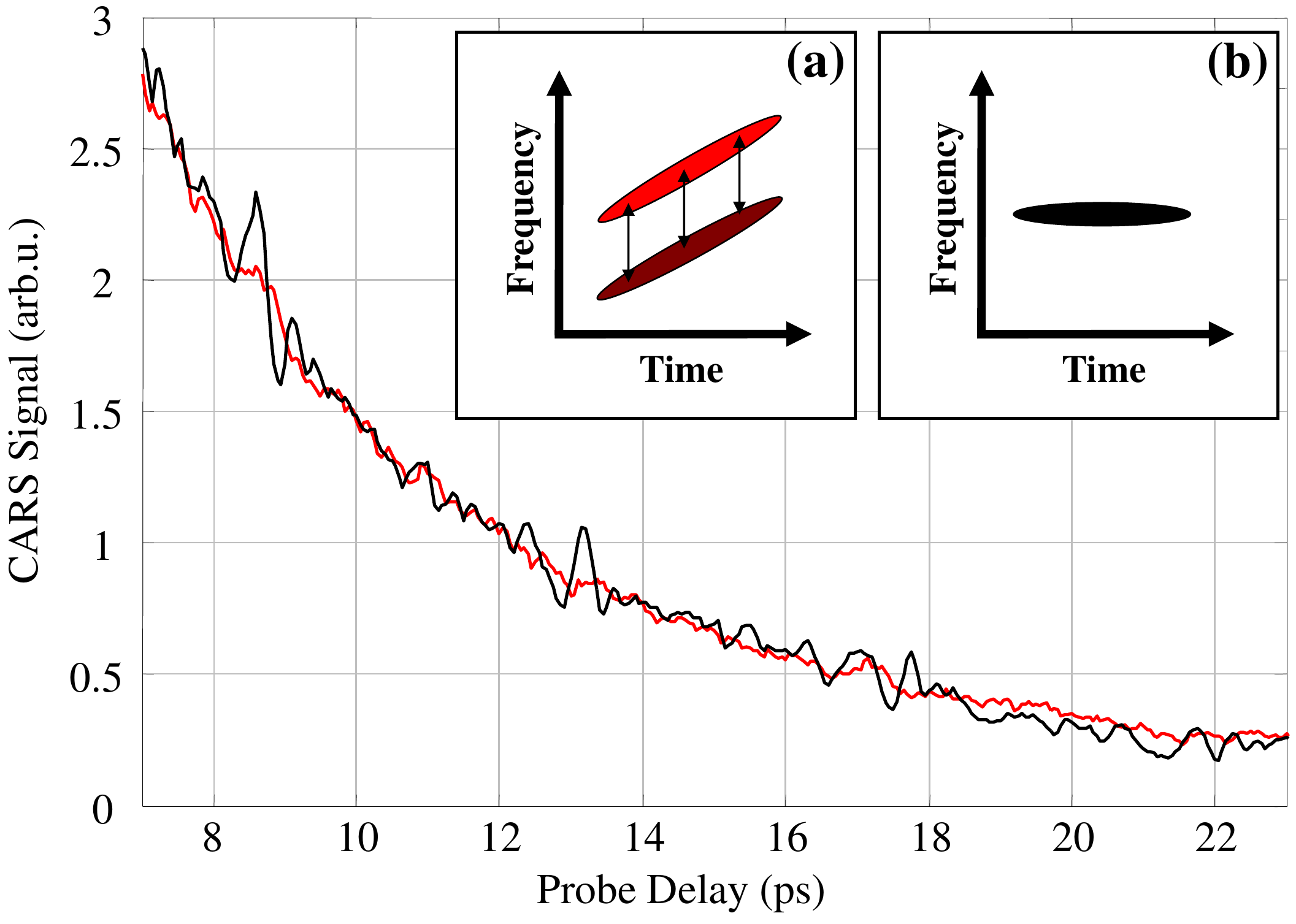}
     \caption{(Color online) Comparison of the dephasing picture for transform-limited pump and Stokes pulses (black) and for pump and Stokes pulses with a positive linear chirp of magnitude $\alpha=35,000$fs$^2$ (red). (a) Husimi plot of the chirped pump and Stokes pulses. (b) Husimi plot of the resulting two-photon spectrum.}
  \vskip -.1truein
  \label{Fig:Chirped}
\end{figure}

To demonstrate the control of vibrational dynamics in a thermal ensemble of rotating molecules, we used the technique of femtosecond pulse shaping to suppress the excitation of $O$ and $S$ branches. This can be achieved by frequency chirping pump and Stokes pulses. If the pulses are chirped in the same direction (Fig.\ref{Fig:Chirped}.a), the resulting two-photon spectrum becomes narrower (Fig.\ref{Fig:Chirped}.b). Adjusting the magnitude of the chirp and the relative timing between the two pulses allowed us to center the narrow two-photon spectrum on the frequency of the $Q$-branch transitions, while suppressing the $O$ and $S$ excitation (see Fig.\ref{Fig:TwoPhotonSpectrum}).

The results are shown in Fig.\ref{Fig:Chirped} where we plot the initial dephasing of the vibrational wave packet for two excitation scenarios. In the case of unshaped transform-limited pump and Stokes pulses, fast rotational revivals around 8.5, 13 and 17 ps are superimposed on top of a monotonic decay of coherence. With the applied frequency chirp, this rotational structure disappears, while the slower ro-vibrational dephasing remains unchanged.

\section{Summary and discussion.}
%-----------------------Summary and Discussion-----------------------------------
This paper presents an experimental and theoretical study of coherent ro-vibrational dynamics in a thermal molecular ensemble. Our scheme involves a femtosecond Raman excitation of vibrational coherence in diatomic nitrogen molecules, and a subsequent probing via coherent anti-Stokes Raman scattering. Experimental data allowed us to determine the second-order ro-vibrational constant for diatomic nitrogen, $\gamma_e=-(2.6\pm0.3)\cdot 10^{-5}$cm$^{-1}$.

The ro-vibrational coupling with a thermal rotational bath leads to a loss of vibrational coherence on a picosecond time scale. At longer times, however, we observe fractional and full revivals of the ro-vibrational wavepacket. The full revival of the vibrational coherence between $v=0$ and $v=1$ was observed at $960.2\pm 0.3$ ps. Notably, the coherent ro-vibrational evolution is observed in a thermal mixture of rotational states rather than in a coherent superposition. In contrast to the well known coherent rotational wave packets, here the dynamics are dictated by the relative phase between the vibrational coherence in molecules occupying different $J$ states in a thermal mixture.

%----------------------Discussion: why we could to it-----------------------------
While the coherent ro-vibrational revivals have been theoretically predicted \cite{Wallentowitz2002}, their experimental observation was considered unlikely. Our experiment is different from the earlier studies \cite{Branderhorst2008,Walmsley1998} in several aspects. First, femtosecond CARS allowed us to excite and probe a two-level vibrational coherence, rather than a many-level vibrational wave packet. Thus the ro-vibrational and rotational dynamics are separated from the vibrational one. Further, pulse shaping allowed us to make a distinction between the coherent rotational and ro-vibrational dynamics. Finally, the energy level structure of diatomic nitrogen is such that only rotations, but not vibrations, are thermally populated at room temperature. The finite number of populated states makes it possible to resolve the complete time evolution of the prepared wave packets. The ro-vibrational revival time can be varied by choosing an appropriate excited vibration level. In addition, the ratio between decoherence due to collisions and the dephasing due to the rotational bath can be easily tuned via temperature and pressure.

%-----------------------Outlook: control of decoherence ---------------------------
The ability to observe and influence molecular evolution with a controlled degree of dephasing, both Markovian (collisions), and non-Markovian (ro-vibrational coupling), offers a good starting point for further investigation toward control of decoherence.

%------------------------Acknowledgement ------------------------------------------
\begin{acknowledgements}
The authors would like to thank Moshe Shapiro, Misha Yu. Ivanov and Ilya Averbukh for valuable discussions.
\end{acknowledgements}

%------------------------Appendix -------------------------------------------------
\section*{Appendix: Branching ratio between the different CARS branches.}
In order to estimate the relative intensity of different Raman transitions, it is convenient to describe them via the coordinate-dependent molecular polarizability. In such a description \cite{Sussman2011}, the electronic coordinate is adiabatically eliminated from the interaction Hamiltonian, whereas the dependance on the nuclear coordinate is left in its explicit form. The molecular Hamiltonian \cite{Friedrich1995,Zon1975} in the combined field of pump and Stokes pulses is expressed as
\begin{eqnarray}
H &=& H_0 - \frac{1}{2} \langle \big( \,E_p \cos \w_p t +
 E_S\cos \w_S  t \,\big)^2 \rangle \, \nonumber \\ %
 && \times\,\left[ \a_\bot(R) + \D\a(R) \cos^2\theta  \right], %
 \label{eq:Raman-full}
\end{eqnarray}
where $E_p$ and $E_S$ are the pump and Stokes field amplitudes, $\w_p$ and $\w_S$ are their frequencies, $\langle  \rangle$ stands for time-averaging over optical oscillations, $\a_\bot(R)$ and $\D\a(R) = \a_\|(R)-\a_\bot(R)$ are the molecular polarizabilities which depend on the internuclear coordinate. Expanding near the
equilibrium internuclear distance $R_e$,
\begin{eqnarray}
\a_\bot(R) &=& \a_{\bot0} + \a_\bot'\,(R-R_e), \nonumber \\ %
\D\a(R) &=& \D\a_0 + \D\a'\,(R-R_e)  ~,
\end{eqnarray}
and time-averaging the squared field term in Eq.(\ref{eq:Raman-full}), gives
\begin{eqnarray}
\label{eq:Ramanmain}
H & = & H_0  \\
%H- & = & H1
 & - & \frac{1}{4} \big( E_p^2 + E_S^2 + 2 E_p E_{S} \cos \W t \big) \, \a_{\bot0}  %
\nonumber \\%
 & - & \frac{1}{4}\big(E_p^2 + E_S^2 + 2 E_p E_{S} \cos \W t \big) \, \a_\bot' (R-R_e) %
\nonumber \\%
& - & \frac{1}{4}\big(E_p^2 + E_S^2 + 2 E_p E_{S}\cos \W t \big) \,\cos^2\theta\, \D\a_0   %
\nonumber \\%
& - & \frac{1}{4}\big(E_p^2 + E_S^2 + 2 E_p E_{S} \cos \W t \big) \,\cos^2\theta\, \D\a' (R-R_e).   %
\nonumber %
\end{eqnarray}
Raman frequency $\W =\w_p-\w_S $ is assumed to be close to the energy difference between the eigenstates $v=1, J'$ and $v=0, J$
for any $J,J'$.

Evaluating the transition matrix elements of various terms in the Hamiltonian (\ref{eq:Ramanmain}), we express the ro-vibrational wave function $|v,J\rangle$ as a product of the $R$-dependent vibrational part and $\theta,\phi$-dependent rotational part. The ro-vibrational matrix elements of the perturbation terms factorize into the products of the vibrational and rotational parts.

The rotational matrix elements of $\cos^2 \theta$ \cite{Zare-book,CB-book} are different for different $M_J$ manifolds. For an estimate, we take their ratio to be that of the case $J\gg M_J$ or, equivalently, of the 2D rotator where
\begin{equation}
\langle J,M_J | \cos^2\theta | J',M_J' \rangle \simeq \left[ %
    \begin{array}{lll}
    1/2,&  J=J',&  M_J=M_J' \\
    1/4,&  J=J'\pm2,& M_J=M_J'%\\
   %0   & \text{otherwise}
    \end{array}
    \right.
\label{eq-cossq}
\end{equation}
The vibrational matrix elements can be evaluated assuming that the molecular potential $V(R)$ is harmonic near the equilibrium. Thus \cite{LL-book}
\begin{eqnarray}
\langle v=0 | R-R_e | v=0 \rangle &=&0, ~\nonumber \\
\langle v=1 | R-R_e | v=1 \rangle &=&0, ~ %
\nonumber \\%
 \langle v=1 | R-R_e | v=0 \rangle &=&
{\sqrt{\hbar / 2 m \w_{01}}},~
 \label{eq-HarmonicOsc}
\end{eqnarray}
where $m$ is the reduced mass of the molecule.

Finally, we apply the Rotating Wave Approximation, and retain only those terms in the Hamiltonian which remain after averaging over the vibrational period. Those terms in Eq.(\ref{eq:Ramanmain}) which contain $\cos \W t$ can have non-vanishing time-averaged matrix elements for transitions between $v=0$ and $v=1$ states, but they can not drive transitions between states with the same value of $v$. The latter transitions are driven by the time-independent terms of the Hamiltonian. After averaging, the Hamiltonian becomes:

\begin{eqnarray}
\label{eq-Raman-final}
H - H_0 &=& \\
 &-& \frac{1}{4}\big(E_p^2 + E_S^2  \big) \, (\cos^2\theta\, \D\a_0 + \a_{\bot0} )   %
\nonumber \\%
 &-& \frac{1}{2} E_p E_S \cos \W t  \, \a_\bot' (R-R_e) %
\nonumber \\%
&-& \frac{1}{2} E_p E_S \cos \W t \,\cos^2\theta\, \D\a' (R-R_e) ~.  %
\nonumber %
\end{eqnarray}
The first term on the right hand side of Eq.(\ref{eq-Raman-final}) is responsible for the AC Stark shift and conventional alignment of $N_2$ by the pump and Stokes laser fields. These effects were not measured in our experiment. The second term drives the transitions between $v=0,J,M_J$ and $v=1,J,M_J$ states. The third term leads both to $\Delta J=0$ and $\Delta J=\pm2$ transitions. The ratio between the $J \leftrightarrow J$ and $J\leftrightarrow J\pm2$ transition amplitudes is due to the ratio of the corresponding matrix elements, and is given by
\begin{equation}\label{eq-ration}
N_{Q/SO} = \frac{ \a_\bot' + \D\a' \langle J |\cos^2 \theta | J
\rangle  }{ \D\a' \langle J\pm 2 |\cos^2 \theta | J \rangle}.
\end{equation}
For nitrogen in an off-resonant laser field, we use the static values of polarizability $\a_\bot' \simeq 8.7 / R_e$, and $\D\a' \simeq 13.3 / R_e$ in atomic units \cite{Jensen1989}. From Eq.(\ref{eq-cossq}) we obtain
\begin{equation}
\label{eq-nsq}
 N_{Q/SO} \simeq 4.6~.
\end{equation}

When the probe field is applied, the Raman transition back to the initial $v,J$ state is governed by a similar two-photon matrix element. The transitions $v=1,J,M_J \rightarrow v=0,J,M_J$ lead to the $Q$-branch radiation, whereas $v=1,J\pm2,M_J \rightarrow v=0,J,M_J$ corresponds to $O$ and $S$ branches. By comparing the matrix elements for populating and probing $v=1,J'=J\pm2$ states with those for $v=1,J'=J$ we see that the amplitudes of the anti-Stokes electric field in $O$ and $S$ branches is $N_{Q/SO}^2\simeq 21$ times weaker than those in $Q$ branch.

%------------------------Bibliography---------------------------------
%\bibliography{Decoherence}

\end{document}